\documentclass[aps,prb,showpacs]{revtex4}
\usepackage{graphicx}
\usepackage{latexsym}
\usepackage{amssymb}
\usepackage{amsmath}
\usepackage{setspace}
\newcommand{\YBCO}{ YBa$_{2}$Cu$_{3}$O$_{7}$ }
\newcommand {\LSMO} { La$_{2 / 3}$Sr$_{1 / 3}$MnO$_{3}$ }
\newcommand{\PBCO} { PrBa$_{2}$Cu$_{3}$O$_{7}$ }
\newcommand{\cuo}{CuO$_2$ }
\newcommand{\sto}{SrTiO$_3$ }
\newcommand{\twotheta}{2$\theta$ }
\newcommand{\onebar}{$(1\overline{1}0)$ }
\begin{document}
\title{Magnetoresistance studies of \LSMO-\YBCO-\LSMO trilayers with ferromagnetic coupling along the nodal direction of \YBCO}

\author{Soumen Mandal\footnote{Present Address: Institut N\'eel, Grenoble, France E-mail: soumen.mandal@gmail.com}}
\affiliation{Department of Physics, Indian Institute of Technology
Kanpur,  Kanpur - 208016, India}


\begin{abstract}
I have successfully prepared (110) trilayers of \LSMO-\YBCO-\LSMO.
Magnetization measurements on these samples reveal a stronger
coupling between the ferromagnetic layers. The coupling is an
order of magnitude higher than that seen in the case of (001)
trilayers. Magnetoresistance measurements show a first order
transition in the data coinciding with the antiferromagnetic
regime deduced from the magnetization measurements. I have also
measured the anisotropic magnetoresistance (AMR) of these samples
revealing an unusually high AMR ($\sim 72000\%$). I attribute such
a high AMR to the pair breaking effects in these films.
\end{abstract}


\maketitle

\section{Introduction}
The study of hybrid structures composed of oxide
superconductor(SC) and oxide ferromagnet(FM) systems have revealed
a variety of exotic phenomena in these systems\cite{Buzdin,
Bergeret, Izyumov, Pokrovski, Demler, Tagirov, Goldman, Chien,
Garifullin, Keizer}. The interest in these systems arises from the
fact that both the states, i.e. SC and FM, are mutually exclusive.
Proximity of such exclusive states can give rise to phenomena like
Larkin - Ovchinnikov - Fulde - Ferrel(LOFF) state, exchange
coupling, Andreev reflection to name a few. Due to recent advances
in thin film fabrication technique it is possible to achieve good
quality interface between SC and FM layers and grow these films in
any desired orientation by use of suitable substrate. It is to be
noted that the same structure in its normal state can act as a
well known spin valve system where the two magnetic layers are
separated by a non-magnetic layer\cite{Dieny, Chaiken}. When the
spacer in such structures is a metallic ferromagnet of 3d and 4f
elements the exchange coupling is chiefly driven by the
Rudderman-Kittel-Kasuya-Yoshida interaction (RKKY)\cite{Bruno}.

In this paper I am chiefly interested in (110) oriented trilayers
of double exchange ferromagnet \LSMO(LSMO) and a non-Fermi-liquid
metal \YBCO (YBCO). The giant magnetoresistance seen in FM-NM-FM
trilayers and multilayers is related to asymmetric scattering of
spin-up and spin-down electrons as they criss-cross the spacer
while diffusing along the plane of the
heterostructure\cite{Binasch, Baibich, Fert}. In case the
intermediate NM layer becomes superconducting one would expect to
see a profound change in the flow of spin polarized carriers. It
is to be noted that most of the earlier studies involving
superconducting spacer layers in a spin-valve type configuration,
where the SC is YBCO, the spin injection in SC layer is along the
insulating c-axis\cite{Goldman, Dong, Pena, Senapati}. Such
structures do not allow the injection of spin polarized carriers
along the fully gapped nodal planes of YBCO. To overcome this
problem it is necessary to grow the films in such a way that the
\cuo planes are perpendicular to the plane of the substrate and in
direct contact with the ferromagnetic layers. This is possible if
the YBCO layer is grown with crystallographic direction
(100)/(010) or (110) perpendicular to the plane of the substrate.
In an earlier work we have already demonstrated the growth of
(110) hybrids\cite{Mandal}. The reason for choosing (110)
orientation over (100)/(010) is explained as follows. The growth
of YBCO films where the c-axis of the film is in the plane of the
substrate involves the use of heterotemplate technique. Now if the
template(\PBCO in this case) is grown (100)/(010) and then the
first LSMO layer is deposited, then the YBCO layer becomes (001)
oriented where the c-axis is perpendicular to the substrate plane.
This is because of the crystallographic symmetry of the LSMO
molecule along (001), (010) and (100) direction. But in the case
of (110) growth the LSMO layer is (110) oriented so the only
possible growth directions for YBCO are (110) and (103). The
template in this case helps in increasing the (110) volume
fraction. In this paper I describe experimental studies of
transport and magnetic properties of nodally coupled hybrids. I
have also carried out some control experiments to demonstrate the
role of ferromagnetic layers on both sides of SC layer.

\section{Experiments}
Thin films of (110) trilayer of LSMO - YBCO - LSMO, bilayer of
YBCO - LSMO and (110) YBCO were deposited on (110) \sto
substrates. A multitarget pulsed laser deposition technique based
on KrF excimer laser($\lambda = 248nm$) was used to deposit the
thin films. The (110) trilayers had 200{\AA} and 500{\AA} YBCO
layer sandwiched between 1000{\AA} of LSMO. The heterostructure
was grown using a heterotemplate technique with the template being
\PBCO. Further details of film deposition giving information about
growth rate, deposition temperature and pressure are given
elsewhere \cite{Mandal}. The epitaxial growth in (110) films were
established by X-ray diffraction measurements performed in
$\theta$ - \twotheta geometry. The volume fraction of (110) grains
in (110) trilayer was determined by the recipe of Westerheim {\it
et al.}\cite{Westerheim} which comes out to be $\geq$65\% with the
remaining volume of (103) grains. It is true that the trilayers do
not have 100\% (110) oriented grains but the presence of (103)
grains still allow direct injection of spin-polarized carriers in
the \cuo planes(Fig. 4.4 from Ref.\cite{thesis}). For
magnetization measurement a commercial magnetometer (Quantum
Design MPMS XL5 SQUID) was used. For transport measurements, films
were patterned in the form of 1000 $\times$ 100 $\mu m^2$ bridge
with photolithography and wet etching such that the long axis of
the bridge was parallel to \onebar direction for the (110)
oriented films. The measurements of resistivity as a function of
temperature, magnetic field strength and the angle$(\theta)$
between field and current were performed using a 4.2K close cycle
He - refrigerator with a fully automated home made setup for
applying the field at varying angles between 0 and 2$\pi$ with
respect to the direction of current\cite{Patnaik}. The sample was
mounted in a way to keep the field in the plane of the sample for
all values of the angle between $\vec{I}$ and $\vec{H}$ except for
the measurements where out of plane contributions were also
recorded.

\section{Results and Discussions}
\subsection{(110) LSMO - YBCO - LSMO}
\begin{figure} \centerline{\includegraphics[width=4in,angle=0]{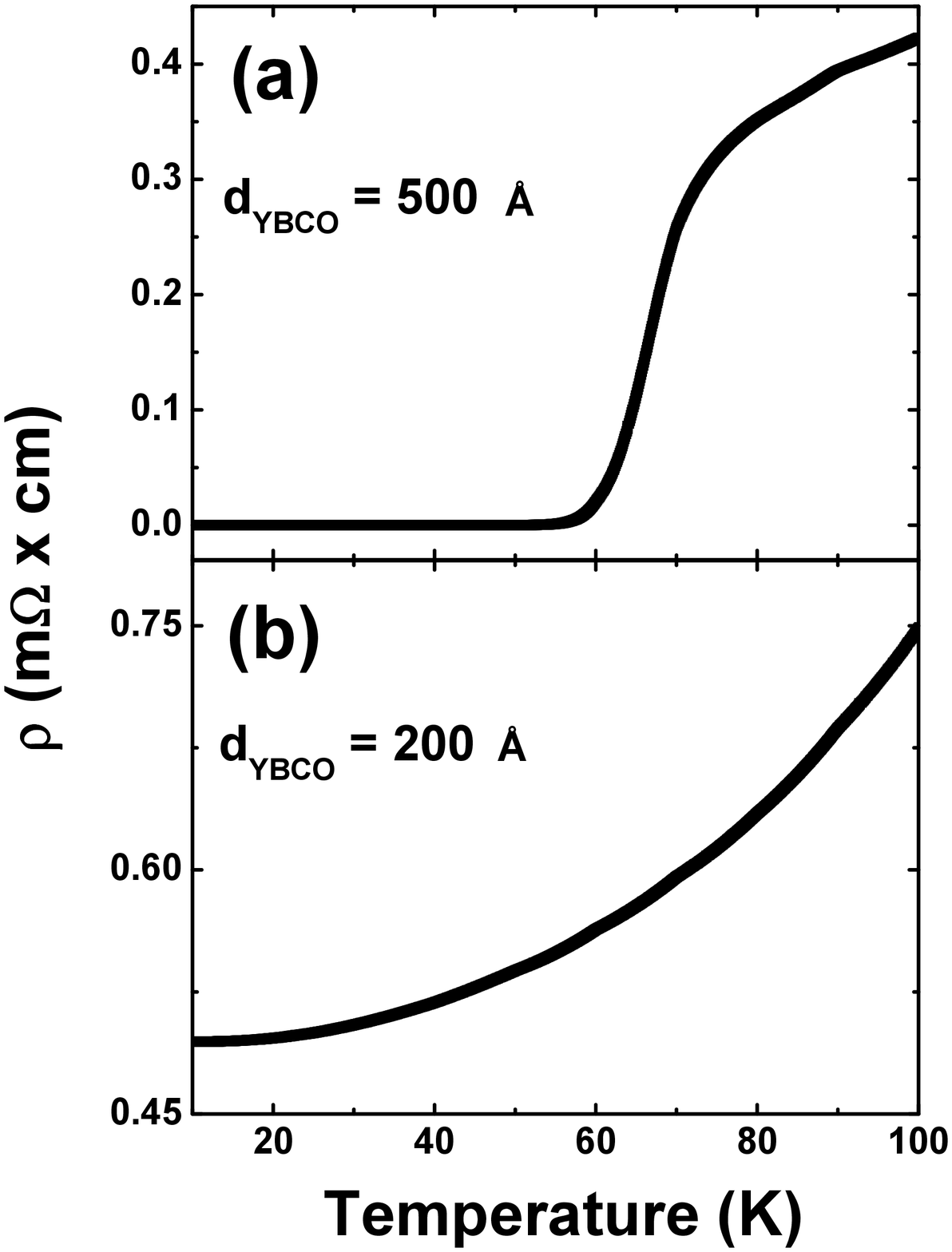}} \caption{Resistivity data for (110) LSMO - YBCO - LSMO trilayer
with d$_{YBCO}$ = 500{\AA} (top panel) and 200{\AA} (bottom
panel). Here the films with 200 {\AA} thick YBCO layer are not
superconducting while (001) trilayer films are superconducting
even for a YBCO layer thickness of 50 {\AA}.}
\label{rtt110}\end{figure}

\begin{figure}\centerline{\includegraphics[width=4.5in,angle=0]{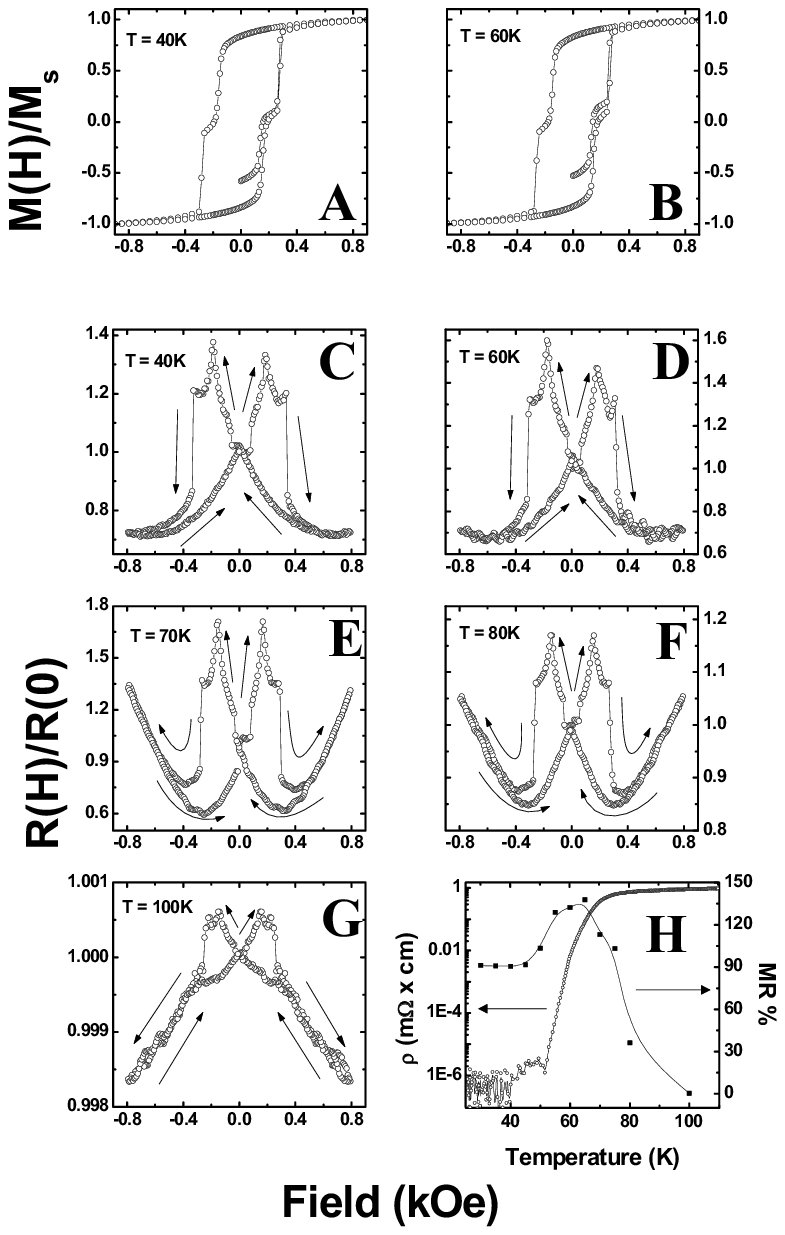}} \caption{Panels A and B show the M-H loop of the trilayer
at 40 K and 60 K respectively. Panels C-G show the field
dependence of resistivity for the superconducting trilayer at a
few representative temperatures across the transition temperature.
The MR of the film in the superconducting state is higher than
that in the normal state. Panel H shows the comparison of MR\%
with temperature and sample resistivity. A clear peak seen near
the transition temperature can be attributed to the unusual rise
in the normal state properties of the superconductor near the
transition temperature.} \label{rht110} \end{figure}

With the optimized growth conditions, trilayers of (110)
\LSMO-\YBCO-\LSMO were synthesized and their various magnetic and
electronic properties were measured. Fig.\ref{rtt110} shows the
resistivity curves, $\rho(T)$, for two trilayers. The upper panel
is the result for a trilayer with a 500 {\AA} YBCO spacer. The
$\rho(T)$ curve is characterized by transition to a
superconducting state which starts at $\sim$80 K and completes
when the temperature reaches $\sim$60 K. The bottom panel shows
the resistivity for a similar structure with a 200 {\AA} YBCO
spacer. In this case the trilayer does not go into the
superconducting state though it has a metallic behavior. It is
interesting to note that while the YBCO of thickness 200 {\AA} in
the (110) trilayer shows no T$_c$, a superconducting transition
can be seen for YBCO thickness of even 50 {\AA} for the (001)
trilayer \cite{Senapati}. This is presumably due to greater T$_c$
suppression in the case of (110) films because of direct injection
of spin polarized carriers in the superconducting \cuo planes of
YBCO. In figs. \ref{rht110}A \& B, I have shown the M-H loops of
the trilayer with d$_{YBCO}$ = 500 {\AA}, at 40 and 60 K
respectively. A plateau in the M-H loop near zero magnetization
confirms the presence of an antiferromagnetic state. This
antiferromagnetic state is present in the normal state of the
superconductor as well. Panels C through G in the same figure show
the magnetoresistance (MR) of the superconducting trilayer at a
few temperatures across the transition. The MR in this case is
defined as R(H)/R(0) where R(H) is the resistance of the sample at
applied field H. The field and current (I) in this case are
coplanar but orthogonal to each other. I first discuss panels C
and D which present the data for the trilayer in the
superconducting state at 40 K and 60 K respectively. Starting from
a fully magnetized state of LSMO layers at 800 Oe the MR first
increases slowly as the field is decreased. At $\sim$400 Oe the
rate of increase becomes faster but remains continuous till the
zero field. On reversing the field, a small step like jump is seen
around $\sim$-50 Oe and then the MR keeps rising to a peak value,
after which, a local minimum is attained followed by a sudden jump
in the MR at $\sim$370 Oe to a much lower value. Further increment
of the field results in a gradual decrease in MR till a reversed
field of 800 Oe is reached. This cycle repeats itself once the
field is decreased from -800 Oe and increased to 800 Oe. I have
measured MR for the sample at a few more temperatures below T$_c$.
In all those measurements I found that the resistance ratio
$(R_{\uparrow\downarrow})_{max}/(R_{\uparrow\uparrow})_{min}$ over
the whole range of measurement is $\sim$2 where
$(R_{\uparrow\downarrow})_{max}$ is the resistance at the peak
position in the MR-H curve and $(R_{\uparrow\uparrow})_{min}$ is
the minimum resistance of the segment of MR-H curve where the
magnetizations of both the FM layers are parallel to each other.
The current flowing through the sample in these measurements is
zero field $\sim$I$_c$ of the sample at that temperature. Panels E
and F show the MR vs. H data for 70 K and 80 K respectively, where
the YBCO layer in the trilayer is in the superconducting
transition region (top panel of fig.\ref{rtt110}). Here one can
see that the high field negative magnetoresistance region, as seen
in panels C and D in the field regime $\sim$ 400 - 800 Oe, is
replaced by a positive magnetoresistance which is completely
opposite to the negative MR seen on LSMO films \cite{Revzin}. The
resistance ratio in panel E is $\sim$3 which is the highest over
the whole range of measurement. This resistance ratio sharply
drops once the film starts entering the normal state as is evident
from the panels showing the MR at 80 K (panel F) and 100 K (panel
G). Panel G shows MR data for 100K, where negative
magnetoresistance is seen in a high field, which is a
characteristic feature of LSMO \cite{Revzin}. Even though in panel
G the resistance ratio is reduced due to the superconducting
spacer entering into the normal state, however the first order
jump in resistance near H$\approx$350 Oe is still present clearly
proving the fact that the resistance ratio is dependent on the
spacer layer properties while the first order transition is
dependent on the FM layer properties. In panel H the MR\% plotted
is defined as $\Delta R / R(0)$ where $\Delta R =
(R_{\uparrow\downarrow})_{max} - (R_{\uparrow\uparrow})_{min}$ and
R(0) is the resistance at zero field at that temperature. A
distinct behavior of MR\% can be seen when the sample becomes
superconducting. The sample in the normal state has a very low MR
but once the sample starts moving into the SC regime, the MR
shoots up rapidly and then comes down to saturate at a constant
value at low temperatures. The increase in MR in the vicinity of
T$_c$ can be attributed to the abnormal increase in the normal
state properties of the superconductor \cite{Mishonov}.

\begin{figure}
\centerline{\includegraphics[width=4in,angle=0]{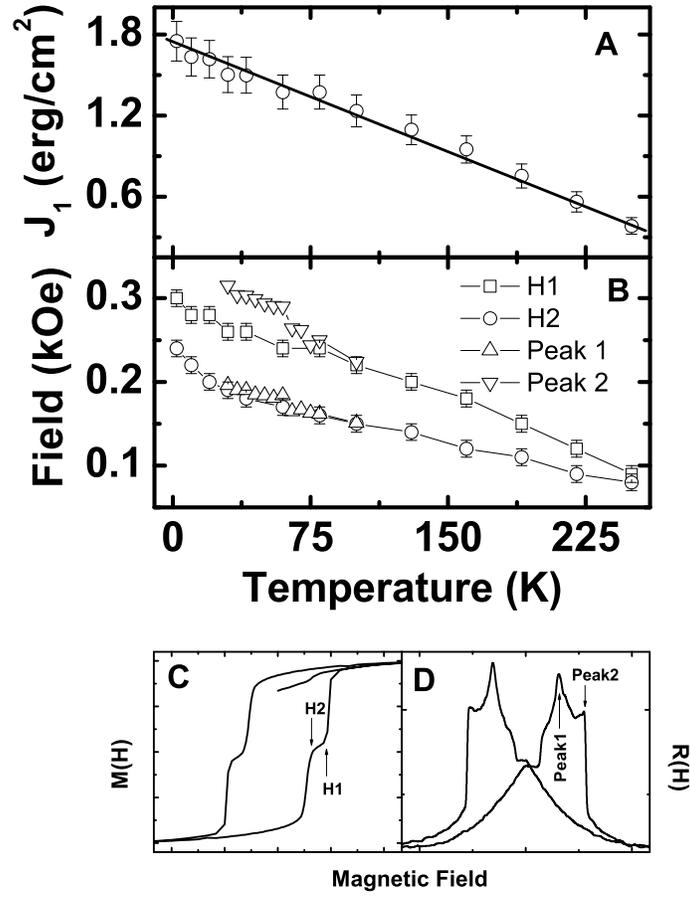}}
\caption{Panel A shows the variation of coupling energy with
temperature between two ferromagnetic layers. The coupling in this
case is higher than that seen for (001) layers (Details in text).
Panel B shows the comparison between the peak position in MR data
and the start and end points of the antiferromagnetic phase in the
MH data. Panels C and D show the position of the points on the MH
and MR data plotted in panel B. The agreement in the data points
clearly shows the dependence of discontinuities in the MR data on
the antiferromagnetic phase of the sample.} \label{j1}
\end{figure}

Another important feature that is quite prominent in
fig.\ref{rht110} is the presence of peaks in the MR data. A
comparison of MR-H and M-H plots (panels A and C, and panels B and
D) shows that the peaks coincide with the region where the M-H
curve has a plateau. The near zero magnetization in the plateau
suggests antiferromagnetic coupling between the magnetization
vectors of the top and bottom LSMO layers. One can estimate the
exchange energy associated with the AF coupling in the following
way. The free energy expression for two magnetic layers of the
same thickness coupled via the spacer can be written as
\cite{Demokritov} \begin{equation} F = F_c + F_a - \vec{H}.
\left(\vec{M_1} + \vec{M_2}\right)t \label{enhs} \end{equation}
where M$_1$ and M$_2$ are the magnetizations of the the top and
bottom LSMO layers, $F_c$ is the coupling energy per unit area and
$t$ is the thickness of a single LSMO layer. The anisotropy part
of the energy ($F_a$) is primarily dependent on contributions from
magnetocrystalline anisotropy as well as the in-plane uniaxial
anisotropy of the film. Assuming a bilinear coupling, $F_c$ can be
written as; \begin{equation} F_c =
-J_1\left(\vec{M_1}.\vec{M_2}\right)\end{equation} where
$\vec{M_1}$ and $\vec{M_2}$ are the unit magnetization vectors,
and J$_1 <$ 0 corresponds to antiferromagnetic coupling between FM
layers. For a given external field, the minima of eq. \ref{enhs}
will yield the relative orientation of $\vec{M_1}$ and
$\vec{M_2}$. If J$_1$ is positive, even in zero field $\vec{M_1}
\| \vec{M_2}$, so increasing the field does not change anything.
If J$_1$ is negative then the minima is achieved when H = 0 and
$\vec{M_1}$ and $\vec{M_2}$ are antiparallel or antiferromagnetic
in alignment. The anisotropic term in eq. \ref{enhs}, F$_a$ can be
written as,
\begin{equation}F_a = KtM_{1,2} \end{equation} where M$_{1,2}$ is a function of $\vec{M_1}$ and $\vec{M_2}$ and $K$ is the anisotropy constant.
If $\mid J_1\mid \gg K$, then a second order reorientation
transition and a smooth linear M-H dependence followed by
saturation is predicted by the theory. On the other hand if $\mid
J_1\mid \ll K$ then the magnetization slowly increases in the low
field, and then abruptly at some critical field H$_s$, the system
undergoes a first order transition with an abrupt jump to
saturation magnetization. The critical field H$_s$ which is also
known as saturation field or switching field can be written in
terms of the magnetization density M$_s$, thickness t of one
ferromagnetic layer and coupling energy J$_1$ as
\cite{Demokritov}, \begin{equation} H_s = -
\left(\frac{J_1}{M_st}\right) \label{eqji}\end{equation} It is to
be noted that this equation is for the case where the two FM
layers are of equal thickness $t$. The behavior of the
magnetization seen in fig.\ref{rht110} corresponds to this
situation. In fig.\ref{j1}(a), I have shown the variation of J$_1$
as a function of the temperature calculated using eq. \ref{eqji}
for d$_{YBCO}$ = 500{\AA}. Panel B of fig.\ref{j1} shows a
comparison between the peak position in the MR-H data and the
start and end points of the antiferromagnetic phase in MH data.
Panels C and D show a typical MH and MR-H data for the sample. The
arrows point to positions of the points on the MR-H and MH data
which have been plotted in panel B.

\begin{figure} \centerline{\includegraphics[width=5in,angle=0]{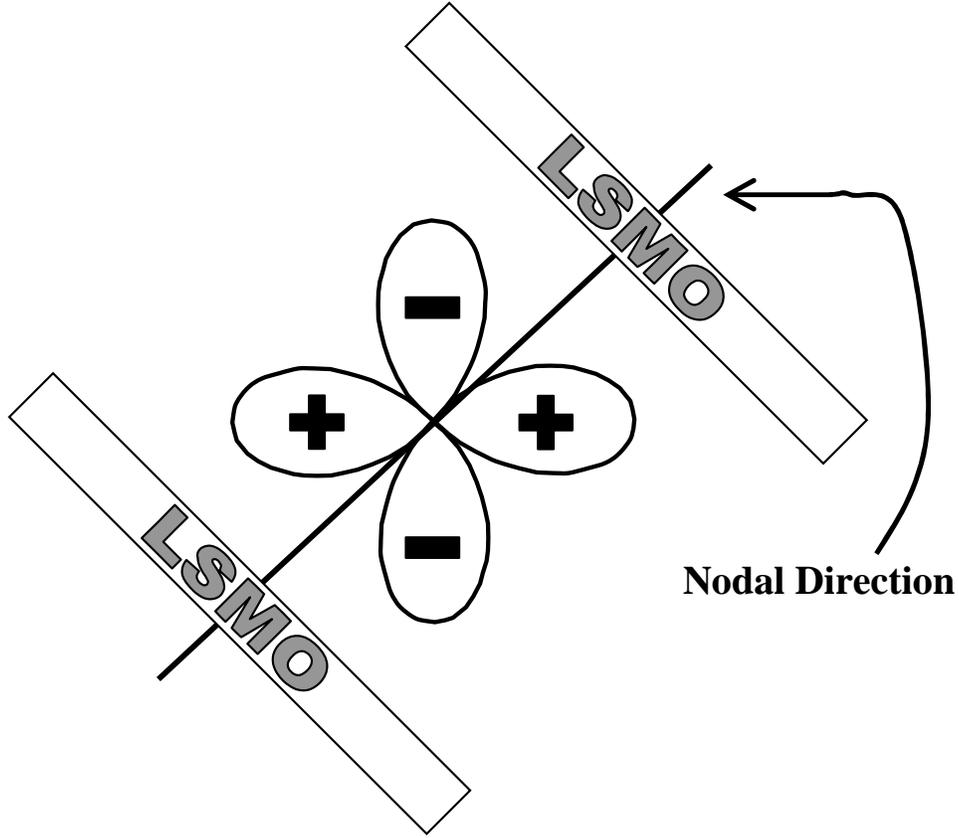}}
\caption{Schematic showing nodal direction in $d_{x^2-y^2}$-orbital. The LSMO layers shown in the schematic signify the position of LSMO layers
in a (110) trilayer. This schematic is for a trilayer where the spacer is a $d$-wave superconductor.} \label{nodaldiro} \end{figure}

I now discuss the behavior of J$_1$ as seen in fig.\ref{j1}(panel
A). The temperature dependence of the interlayer exchange coupling
in metallic multilayers has been worked out theoretically
\cite{Edwards, Brunoprb}. The starting point for calculating the
coupling is to calculate the energy per unit area in the
ferromagnetic and antiferromagnetic configuration. The difference
of the two will give the exchange coupling of the system. The
energy terms are functions of the reflection coefficients of the
electrons in the spacer hitting the spacer-ferromagnet interface
calculated in the light of the free-electron model. Using the
above method the dependence of linear exchange coupling J$_1$ with
temperature is given by \cite{Brunoprb} \begin{equation} J_1(T) =
J_1(0)\left[\frac{\left(T/T_0\right)}{\sinh\left(T/T_0\right)}\right]
\label{j1t} \end{equation} where the characteristic temperature
$T_0$ depends on the Fermi wave vector $k_F$ and spacer thickness
$d_n$ through the relation $T_0 = \hbar^2k_F/2\pi k_Bd_nm$, where
$m$ is the free-electron mass and $\hbar$ and $k_B$ are the Planck
and Boltzmann constants respectively. In this case, since the
transport is along the (110) axis, the relevant wave vector will
be $k_{F_{(110)}}$. In fig.\ref{j1} it is seen that $J_1$
increases linearly as the temperature decreases. This is different
from the behavior expected from eq. \ref{j1t}. In general,
functions of the type x/sinh(x) saturate in the limit x
$\rightarrow$ 0, but in this case I do not see any saturation of
J$_1$ even at a temperature of 2 K. The magnitude of $J_1$ is
almost an order higher than what is seen for (001) oriented
heterostructures \cite{Senapati}. These high values are in line
with the predictions of de Melo \cite{Melo} where he had pointed
out that the coupling along the (110) direction will be higher
than that along the (100) or (001) directions. In
fig.\ref{nodaldiro}, I have shown the schematic of a (110)
trilayer where the spacer is a $d$-wave superconductor. It is
clear from the figure that in this case the coupling, is mediated
by the nodal quasiparticles whose number density remains high even
at T$\approx$0. This explains the large J$_1$ and the absence of
any anomaly in J$_1$ near T$_c$. The middle panel of fig.\ref{j1}
shows the comparison between various critical points on the MH
loop and MR data. Here I have plotted the starting and end points
of the plateau in the MH loop against temperature. From the MR(H)
loop, I have taken the points of discontinuity which are indicated
in panel D. Panel C shows the position of points H1 and H2 on the
MH loop. One can clearly see that both the representative points
agree well within experimental error with each other clearly
demonstrating the fact that the discontinuities in MR data come
from the antiferromagnetic regime of the sample.

\begin{figure} \centerline{\includegraphics[height=5in,angle=-90]{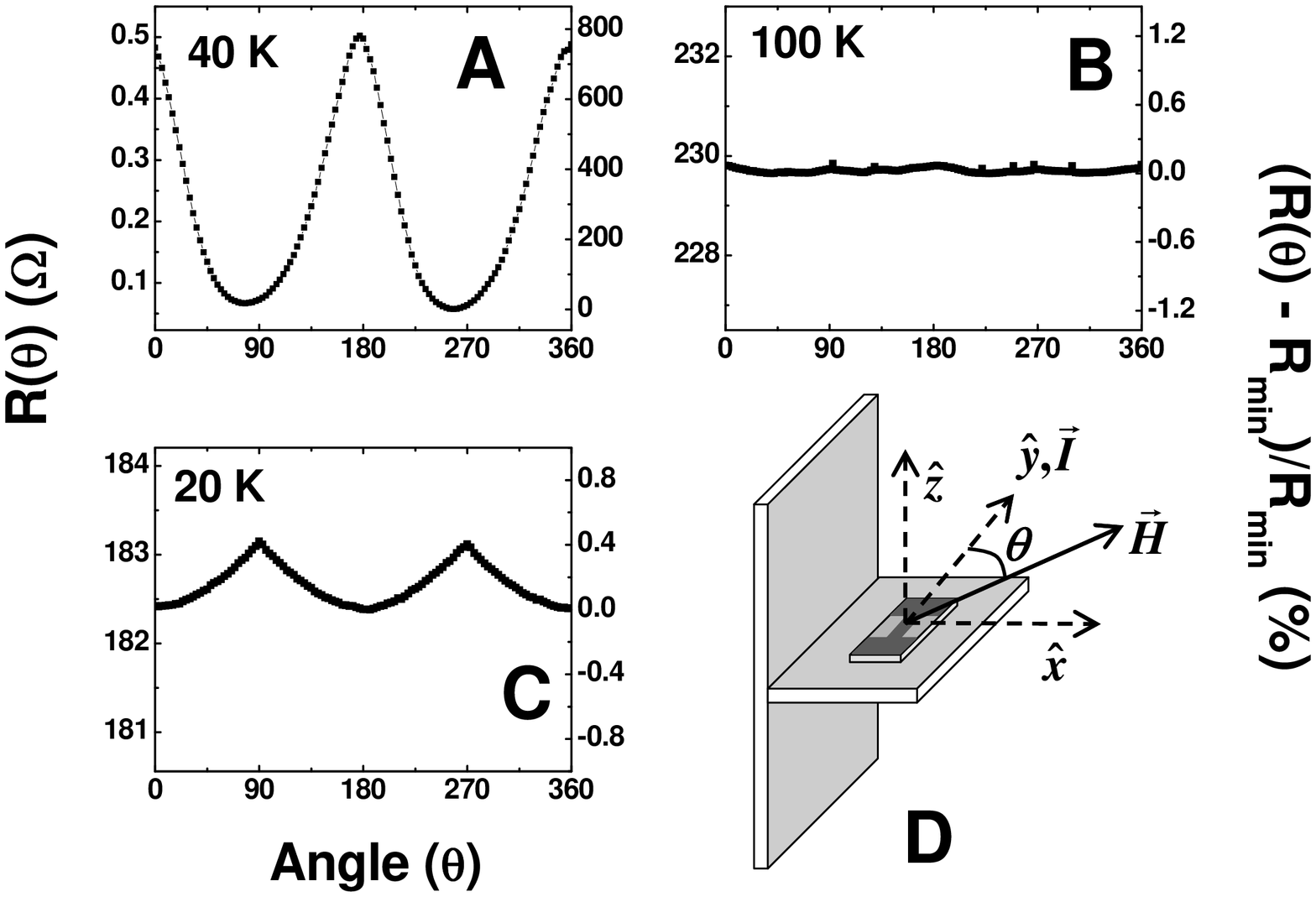}}
\caption{Angular dependence of MR in the FM-SC-FM trilayer is
plotted for T $<$ T$_c$ (panel A), T $>$ T$_c$ (panel B) and a
FM-NM-FM trilayer (panel C). The data clearly shows that a
superconducting spacer in the superconducting state enhances and
modifies the AMR considerably. The AMR of the FM-NM-FM trilayer is
mostly dependent on the AMR of the FM layer. Panel D shows the
schematic of the measurement geometry.} \label{tcom}\end{figure}

\begin{figure} \centerline{\includegraphics[height=6.5in,angle=0]{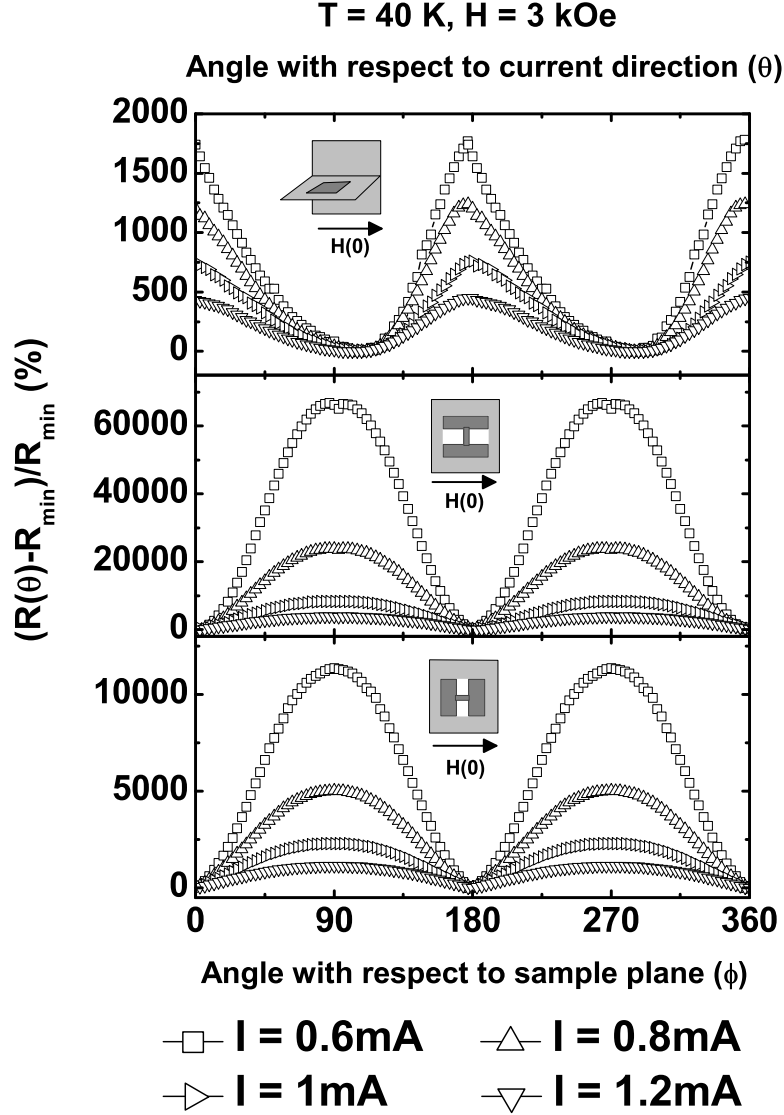}}
\caption{Current dependence of AMR measured at three different
configurations. The top panel shows the AMR for the configuration
when the applied magnetic field stays in the plane of the sample.
The middle panel shows the AMR when the current through the sample
is always perpendicular to the field. The field in this case moves
from a position where it is parallel to \cuo planes to a position
where it is perpendicular to \cuo planes. The bottom panel shows
the AMR when the applied field moves in the plane of the \cuo
planes. The AMR of the sample increases as the current through the
sample is decreased, clearly proving the fact that a low
resistance spacer enhances MR in these trilayers.} \label{amri}
\end{figure}

Fig.\ref{tcom} presents the AMR data on (110) trilayers. The
x-axis defines the angle $\theta$ with respect to the current
direction. The left hand axis shows R$(\theta)$ and the right hand
y-axis shows MR \% defined as (R$(\theta)$ - R$_{min}$)/R$_{min}$
where R$_{min}$ is the minimum resistance of the sample over the
whole range of measurement. The reason for using this definition
will be explained when we discuss the data presented in fig.
\ref{amri}. Before I discuss the data in detail, let me explain
the measurement geometry which has been schematically shown in
panel D of the diagram. The patterned sample is placed over a
solid block of copper as shown in the figure. The current in the
sample is along \emph{\^y}. The applied field rotates in the
$xy$-plane. The angle $\theta$ is measured with respect to
\emph{\^y}. The sample is patterned along \emph{\^y} in such a way
that the \cuo planes of YBCO are in the $yz$-plane. In short,
\emph{\^x, \^y} and \emph{\^z} are parallel to (001), {\onebar}
and (110) directions of the sample. Panels A and B show the AMR of
the trilayer at T = 40K ($<$ T$_c$) and T = 100 K ($>$ T$_c$)
respectively on a film with d$_{YBCO}$ = 500 {\AA}. Panel C shows
the AMR measurement at 20 K on a film with d$_{YBCO}$ = 200 {\AA}.
From panels A \& B, one can see that the trilayer shows a huge MR
when the SC layer is in the superconducting state. The same film
hardly shows any AMR once the film moves into the normal state.
This is also evident from the AMR data in panel C. The trilayer in
this case is non superconducting for all temperatures (bottom
panel of fig.\ref{rtt110}). The angular dependence in panel C is
similar to the one seen for plain LSMO films\cite{Mandal1}. The
dependence of AMR in the superconducting state is markedly
different from the one in the normal state. For my sample
geometry, when the field is perpendicular to the current it is
also perpendicular to the \cuo planes resulting in maximum
dissipation in the YBCO layer. So, the logical thing would be that
the AMR is higher when the field is perpendicular to the current,
but what I see here is completely opposite. This can be explained
as follows. We know that the dissipation in YBCO when the applied
field is perpendicular to the copper oxide planes is due to the
formation of vortices and when the field is parallel, the
dissipation is mostly due to pair-breaking effects. In my
geometry, the effective area of the sample exposed perpendicular
and parallel to field is equal to the thickness of the film times
the length and breadth respectively. It is quite possible that the
effect of vortex formation in such a small area has lower
dissipation than pair breaking effects. Hence, if I assume that
pair breaking is causing larger dissipation in the YBCO layer in
these trilayers, I can safely conclude that for fields parallel to
the current (or copper oxide planes) the AMR will be higher.

In fig.\ref{amri}, I have plotted the current dependence of the
AMR for three different orientations of the sample as shown in the
figure. Before I explain the results in this figure it is
important to discuss the reasons behind using the particular
definition for AMR. If we look closely H(0) in the figure
corresponds to the situation when the angle of the field with
respect to the current or sample plane is zero. In the top panel,
the measurement geometry is the same as is shown in
fig.\ref{tcom}(d). In the middle panel, the \cuo planes and
current are perpendicular to field when $\phi = 0$ and in the
bottom panel the field is parallel to the \cuo planes and current
for H(0). So, to compare the data in these three panels it is
important to find a equivalence point for the calculation of AMR
which one cannot find if one considers the field orientation or
\cuo plane orientation or current as reference point. So the best
option would be to consider the point where the resistance of the
sample is minimum. Coming back to the top panel, which shows the
dependence in the coplanar configuration, one can see that as
current through the sample is decreased AMR increases. For the
same film, the middle and bottom panel show unusually large AMR.
It is to be noted that the AMR in the middle and bottom panel come
due to contributions from two different effects. First will be the
AMR due to field being parallel or perpendicular to the \cuo layer
of the SC spacer and the second will be due to the out-of-plane
field which gives rise to a high resistance state. In the bottom
panel, one can see that the applied field always stays parallel to
the \cuo planes. The AMR seen here essentially arises due to the
fact that the applied field becomes perpendicular to SC layer. In
the middle panel, one can see that the AMR is much higher than
that of any other configuration. In case I assume that the
contribution is only from the first case, as pointed out earlier,
then there should be no angular dependence in the bottom panel and
if I assume that all the contribution is coming from the second
case, then the middle and bottom panels should show comparable
AMR. The fact that the middle panel shows an AMR almost 6 times
that of the bottom panel points to the fact that the AMR is coming
from both the contributions which in this case, have same behavior
in the configuration shown in middle panel and hence the effect is
additive. One can also see that in all the cases, reduction in the
current results in an increase in the AMR vindicating the
hypothesis that low resistance in the spacer contributes to a
higher magnetoresistance. Apart from this, if we try to talk only
about the magnitude of change in resistance and not the percentage
then we will see that the maximum resistance over the whole range
of 2$\pi$ can increase by as much as $\sim$700 times the minimum
resistance. The discussion on the trilayers will not be complete
without a small discussion on the crystallographic differences
along the two directions in the plane of the film. The directions
in the plane of the film are \onebar and (001) with (001) being
the easy axis. The easy axis is predominantly determined by the
easy axis of the LSMO layers which in this case is along the (001)
direction\cite{Mandal1}. When the field is along the (001)
direction it is the easy axis of the LSMO layers but the
dissipative state of the YBCO layer and when the field is along
the \onebar direction it is parallel to \cuo layers which is less
dissipative but the presence of LSMO layer introduces a strong
dissipation resulting in a high resistance state as seen in
figures \ref{tcom} and \ref{amri}.

\subsection{Anisotropic Magnetoresistance of (110) LSMO - YBCO and (110) YBCO thin films.}
\begin{figure} \centerline{\includegraphics[height=7.5in,angle=0]{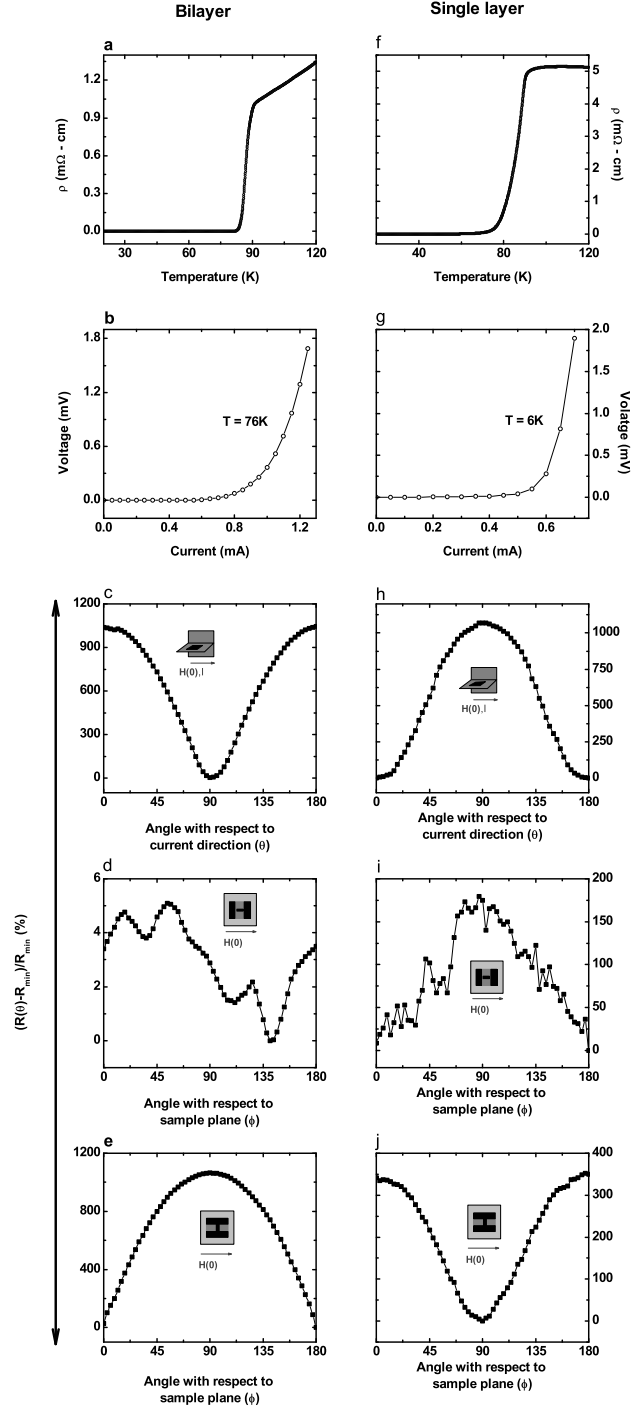}}
\caption{Panel (a) shows the resistivity data for a (110) LSMO -
YBCO bilayer. Panel (b) shows the current voltage characteristic
for the bilayer film at 76 K. Panels (c) to (e) show the AMR for a
LSMO-YBCO bilayer in three different configurations. One can see
that the AMR in this case is almost two orders of magnitude
smaller than that seen for the trilayer. Panel (f) shows the
resistivity data for a (110) oriented YBCO layer. Panel (g) shows
the current voltage characteristic for the YBCO film at 6 K.
Panels (h) to (j) show the AMR for (110) YBCO film in three
different configurations. The AMR primarily arises from the
orientation of magnetic field with respect to \cuo planes.}
\label{rtivbi}
\end{figure}

To verify my results of unusually high AMR in the trilayers I have
done some control experiments involving an LSMO-YBCO bilayer and a
YBCO single layer film. The growth conditions for these films are
exactly the same as the trilayer except for the fact that these
films were made with d$_{YBCO}$ = 1000{\AA}. Fig.\ref{rtivbi}a
shows the resistivity data for the bilayer. Panel b in the same
figure shows a current voltage characteristic for the bilayer at
76 K. In fig.\ref{rtivbi}c, d \& e, I have shown the AMR in this
film at 3 kOe for three different configurations. One can see that
AMR in this film is much smaller as compared to that of the
superconducting trilayer. Most of the contribution comes from the
fact that the field moves from parallel to perpendicular to the
\cuo planes. Panel d shows the AMR when the field stays in the
\cuo plane. The dependence seen here mostly comes from the fact
that the field moves in and out of plane of the sample. In
fig.\ref{rtivbi}f, I have plotted the resistivity data for a
single layer of (110) YBCO. Panel g shows the current voltage
characteristic for the film at 6 K. Fig.\ref{rtivbi}h, i \& j show
the AMR for this film at different configuration. Here one can
clearly see that angular dependence comes primarily from the
positioning of the field perpendicular or parallel to the \cuo
planes. But a look at panels e and i of the figure tells us that
the AMR for the bilayer is higher than that for the single layer.
This is explained by the presence of an FM layer near a
superconducting layer. Earlier, people had seen a higher change in
the resistivity in a bilayer than on a single layer when the film
was pushed into the superconducting state \cite{Petrashov}.
\section{Conclusion}
In conclusion, manganite - cuprate bilayers and trilayers where
the \cuo planes are normal to the plane of the templated (110)
\sto have been synthesized and their various transport and
magnetic properties have been studied. I find that the coupling
between the two FM layers is higher in this case than on that of
the (001) bilayer as predicted by de Melo \cite{Melo}. I have also
observed unusually high ($\sim72000\%$) angular magnetoresistance
in these trilayers. Some control experiments have been done to
point out the fact that the unusually high AMR comes from the
coupling between the two ferromagnetic layers. The MR\% calculated
from the magnetoresistance measurements shows a peak near the
superconducting transition temperature which has been attributed
to the unusual increase in the normal state properties of the
superconductor near its transition temperature.

\section*{Acknowledgements}
The author would like to acknowledge financial support from Indian
Institute of Technology Kanpur, Kanpur, India.

\begin{singlespace}
\newpage

\newpage

\newpage

\newpage

\newpage

\newpage

\newpage

\end{singlespace}
\end{document}